\newcommand \lp{\lambda_{\perp}}
\newcommand \K{{\rm K}_0}
\documentstyle[aps,twocolumn]{revtex}
\begin{document}
\twocolumn[\hsize\textwidth\columnwidth\hsize\csname      
@twocolumnfalse\endcsname                                 

\title{Quantum Melting of a Two-Dimensional Vortex Lattice at Zero Temperature}

\author{A. Rozhkov and D. Stroud}

\address{Department of Physics, Ohio State University, Columbus, OH 43210}

\date{\today}

\maketitle

\begin{abstract}

We consider the quantum melting of a two-dimensional flux lattice 
at temperature T = 0 in the ``superclean limit.''  
In this regime, we find that vortex motion is dominated by the Magnus force.
A Lindemann criterion predicts melting when
$n_v/n_p \geq~\beta$, where $n_v$ and 
$n_p$ are the areal number densities of vortex
pancakes and Cooper pairs, and $\beta \approx 0.1$.
A second criterion is
derived by using Wigner crystal and Laughlin wave functions 
for the solid and liquid phases
respectively, and setting the two energies equal.  
This gives a melting value similar to the Lindemann result.
We discuss the numerical value of the melting field at $T = 0$ for thin
layers of low-T$_c$ superconductor, such as $a-MoGe$, and single layers of
high-T$_c$ materials.


\end{abstract}

\vskip2pc] \narrowtext    

\section{Introduction.}

Vortices in the layered high-T$_c$ materials have remarkably strong
thermal fluctuations, which have been extensively studied\cite{review}.
At sufficiently low temperatures, vortex lines are also expected to be 
subject to \underline{quantum} fluctuations.  Quantum effects should manifest 
themselves in the zero-point motion of vortex lines. If these are large 
enough, the flux lattice can melt even at temperature T=0.
Indeed, many experiments
suggest that vortex lattice melting, both in high-T$_c$
materials\cite{andrade,schilling,fukuzumi} and 
in low-T$_c$ films and multilayers\cite{attanasio}, 
is strongly influenced by quantum fluctuations.  

Several authors have already considered the possibility of quantum
melting in the high-T$_c$ superconductors.   Blatter and
Ivlev\cite{blatter1} have examined the influence of quantum fluctuations
at finite temperatures.   They estimated the shift in the melting curve using a
Lindemann criterion, assuming overdamped dynamics combined with a
Matsubara formalism.   Chudnovsky\cite{chudnovsky} has studied
a hypothetical two-dimensional (2D) 
quantum vortex liquid state at temperature $T = 0$.
Onogi and Doniach\cite{onogi} computed the $T=0$ melting field for a
2D superconductor using  quantum Monte Carlo (QMC) techniques without
dissipative quantum tunneling. 
By taking into account a fictitious magnetic
field arising from the Magnus force 
on the vortex pancakes\cite{ao}, they also found strong numerical evidence 
for fractional quantum Hall (FQHE) states in the vortex liquid.  
Such FQHE states 
had been predicted by several authors\cite{choi,stern}, but principally in the
context of Josephson junction arrays.

In this paper, we describe two simple, quasi-analytical models for estimating
the conditions for quantum melting of a 2D vortex lattice at T = 0, in which 
the fictitious magnetic field is explicitly included.  The same model might
also apply to one layer in a 3D stack of uncoupled layers of high-T$_c$
material.  The first estimate is a simple Lindemann criterion.
The second involves a simple comparison of internal energies in the crystalline
and liquid phases.

\section{Fictitious Magnetic Field and Lindemann Melting Criterion.}

In our model, the vortex pancakes experience
two types of forces: those due to other pancakes, and the Magnus force
arising from the density of Cooper pairs.  
We neglect dissipative forces from the ``viscous'' normal electron
background, as may be acceptable in the
so-called superclean limit\cite{blatter95}.
Of the two remaining forces, the Magnus force usually dominates (see below).
The melting Lindemann melting criterion proves \underline{independent}
of the vortex mass.  By contrast,
in the opposite limit where the intervortex forces dominate, the melting
field depends sensitively on the vortex mass\cite{onogi}.

The so-called Magnus force\cite{ao} 
acting on a single two-dimensional (``pancake'') vortex, in its rest frame,
is
\begin{equation}
{\bf F}_p = q_v h{\bf v} \times {\bf \hat{z}}n_p \equiv \frac{2e}{c}{\bf v}
\times {\bf \hat{z}}B_{eff}.
\end{equation}
Here $q_v = \pm 1$ is the effective charge of the pancake vortex, 
$h$ is Planck's constant, ${\bf v}$ is the pancake velocity, 
n$_p$ is the effective areal number density of Cooper pairs 
(discussed further below), $B_{eff} = \Phi_0n_p$ is the fictitious field,
$\Phi_0 = hc/2e$ is the flux quantum,
and the film is assumed perpendicular to the $z$-axis.

We now wish to show that the intervortex force is typically
small compared to the Magnus force.  For simplicity, consider a
superconducting film of thickness $d$ and London penetration depth $\lambda$.
The direct interaction potential between two pancakes separated by $r$
is
\begin{equation}
\Pi(r) = 2\epsilon_0 d \K \left({r\over\lambda_{\perp}}\right),
\end{equation}
where $\lambda_{\perp} = \lambda^2/d$ is the transverse penetration depth,
$\epsilon_0=\Phi_{0}^2/(16\pi^2\lambda^2)$
gives the energy scale of the interaction per unit length, 
and $K_0(x)$ is the modified Bessel function of zeroth order.  
To estimate the effects of the vortex-vortex
interaction, we assume that the vortices are ordered into a
triangular lattice, and calculate the change in potential energy per vortex, 
$\Delta U_{harm}$, due to
harmonic vibrations about this lattice.  After some algebra, 
this extra energy is found to take the form
$\Delta U_{harm} =\sum_{\bf l} {\chi(l)\over 4}\langle|{\bf u}_0 -
{\bf u}_{\bf l}|^2\rangle$.
Here ${\bf l}$ is a lattice vector of the triangular lattice,
${\bf u}_{\bf l}$ is the displacement of the ${\bf l}^{th}$ vortex from 
equilibrium, and 
$\chi(l)=\frac{\epsilon_0 d}{\lp^2}\left[{\lp\over l}\K^{'}
\left({l\over \lp}\right)+\K^{''}\left({l\over\lp}\right)\right]$,
where $\ell = |{\bf \ell}|$, and the primes denote 
differentiation.

We estimate this energy as follows.
First, since the 
vortex-vortex interaction is assumed small, we neglect 
$\langle {\bf u_0}\cdot {\bf u}_{\bf l}\rangle$.   
Secondly, 
in the weak-screening regime where the nearest neighbor intervortex
distance $a_0<<\lp$, the summation may reasonably be replaced by
an integral. With these approximations, and using several identities for
derivatives of Bessel functions, we finally obtain
\begin{equation}
\Delta U_{harm} \approx (\epsilon_0 d)\times\pi n_v
\langle|{\bf u}_0|^2\rangle,
\end{equation}
where $n_v = 2/(\sqrt{3}a_0^2)$ is the areal vortex density.

Similarly, for a pancake of mass $m_v$ moving in a fictitious
field $B_{eff}$, the zero-point energy per pancake $\Delta U_{mag}$
for a pancake in the lowest Landau level is
\begin{equation}
\Delta U_{mag} = \frac{1}{2}\hbar\omega_c^{eff}, 
\end{equation}
where $\omega_c^{eff} = 2eB_{eff}/(m_vc)$.  

To show that the zero point motion is usually dominated by $B_{eff}$.
we will demonstrate that 
$\hbar\omega_c \ll \hbar \omega_c^{eff}$, 
where $\omega_c$ is the frequency for zero-point motion 
of the harmonic lattice in the absence of B$_{eff}$.
Now $\omega_c = \sqrt{k/m_v}$, where
$k$ is the effective spring constant of the
harmonic lattice.  It follows from eq.\ (3) that 
$k = 2\epsilon_0d\pi n_v$.  

To compare $\omega_c$ and $\omega_c^{eff}$,
we use the London estimate for the penetration depth 
$\lambda^2(T)=(m_p c^2)/(4\pi q^2 n_p^{3D})=(m_p c^2 d)/(4 \pi q^2 n_p)$,
where $n_p^{3D}$ is the pair density
per unit volume, $m_p$ is the pair mass, and $q$ the pair charge.  
Then a little algebra reveals that $\omega_c \ll \omega_c^{eff}$
provided that
\begin{equation}
\frac{m_v}{m_p} \ll \frac{2n_p}{n_v},
\end{equation}
where $m_p$ is the Cooper pair mass .
As will be shown below, $n_v /n_p \approx 0.1$ at the melting
point.  Then inequality (9) is satisfied so long as
$m_v/m_p \ll 20$.  Now in BiSr$_2$Ca$_2$Cu$_2$O$_{8+x}$, the
mass of a single pancake vortex, assuming a thickness 
$d \approx 10 \AA$ (appropriate for a single layer of high-T$_c$ material)
has been estimated as one electron mass\cite{onogi}.  
Thus, in this regime, the inequality is satisfied and 
$\Delta U_{harm} \ll \Delta U_{mag}$ as required.  
Hence, in calculating melting behavior for vortices of this mass, 
we apparently need consider only $\Delta U_{mag}$.  
Our results based on considering only $\Delta U_{mag}$
do indeed give $n_v/n_p \approx 0.1$, thereby confirming the self-consistency 
of our approach.

We now obtain a simple Lindemann melting criterion, assuming
that the dominant contribution to zero-point vortex motion
arises from $B_{eff}$. 
Although $\omega_c^{eff}$ clearly
depends on $m_v$, the corresponding zero-point displacement does not. 
We calculate this displacement assuming the symmetric gauge for the
fictitious vector potential, 
${\bf A}_{eff} = \frac{1}{2}{\bf B}_{eff} \times {\bf r}$.  
Then in the lowest Landau level, one finds
\begin{equation}
\langle |{\bf u}_0|^2\rangle \equiv \langle (u_x^2 + u_y^2) \rangle = 
\frac{\Phi_0}{\pi B_{eff}}= {1\over{\pi n_p}},
\end{equation}
\underline{independent} of vortex mass.

According to the Lindemann criterion, melting occurs
the zero-point amplitude is a certain fraction, say
$\alpha_L$, of $a_0$.
In most conventional materials, $\alpha_L \approx 0.1-0.2$.  
Since $a_0 = (2\Phi_0/\sqrt{3}B)^{1/2}$, the Lindemann
criterion becomes
\begin{equation}
\frac{n_v}{n_p} = \frac{2\pi}{\sqrt{3}}\alpha_L^2 \approx 0.07,
\end{equation}
using the estimate $\alpha_L^2 \approx 0.02$.  
Thus, the Lindemann picture predicts quantum melting 
at T = 0 at a vortex density of around 7\% of the effective
density of Cooper pairs per 
layer.  

\section{Laughlin Liquid Versus Wigner Crystal.}

Next, we describe an alternative way of estimating the melting temperature
in a 2D lattice.   We treat the pancake vortices 
as bosons, moving in the effective magnetic field $B_{eff}$.
To describe the bosons, we use a Wigner crystal (WC) 
wave function in the
solid phase, and a properly symmetrized Laughlin wave function 
in the liquid state.  The melting point is determined by the condition that
the energies $E_{WC}$ and $E_{LL}$ 
of the solid and liquid states should be equal.  A related approach has
been used to discuss melting of the 2D electron lattice in a magnetic 
field\cite{maki,lam} 

The WC wave function may be written 
\begin{equation}
\Psi_{WC}=A\ {\cal S}\left(\prod_{\bf l}\psi({\bf r_l}-{\bf l})\right).
\end{equation}
Here $\psi({\bf r})$ denotes the zero-momentum single-particle
wave-function of the lowest Landau level,
${\cal S}$ is the symmetrization operator, and $A$ is a normalization constant.
We wish to calculate
the averaged vortex-vortex interaction energy in this state, i.\ e.\
$E_{WC}/(2\epsilon_0dS)
=\langle\Psi|\sum_{\bf l_1}\sum_{\bf l_2\neq l_1}\K\Bigl({{|
{\bf r_{l_1}}-{\bf r_{l_2}}|}\over\lp}\Bigr)|\Psi\rangle/2S$,
where $S$ is the sample surface area.

We simplify the calculation by several approximations.  First, since
$a_0 ^2 \gg \langle |{\bf u}_0|^2\rangle$, 
the wave function symmetrization is quantitatively unimportant
for calculating $E_{WC}$. 
Indeed, for large argument, the single-particle
wave function $\psi({\bf r})$
decays exponentially, and the overlap integral between 
$\psi({\bf r}-{\bf l}_1)$ and $\psi({\bf r}-{\bf l}_2)$ is almost zero, unless
${\bf l}_1={\bf l}_2$.  In view of this degree of localization, 
$E_{WC}$ can be expanded in powers of the small ratio
$\langle |{\bf u}_0|^2\rangle/\lp^2$, keeping only the first
two terms.  The result is
$E_{WC}/(2\epsilon_0 dS)=(n_v/2)\sum_{\bf l \neq 0}
\K\Bigl({l\over\lp}\Bigr)
+n_v{\Delta U_{harm}\over{2\epsilon_0 d}}$,
where $\Delta U_{harm}$ is given by eq.\ (3).
The fluctuations $\langle |{\bf u}_0|^2\rangle$ appearing in eq.\ (3)
are, as noted previously, the sum of two parts: 
one due to $B_{eff}$ and
the other to the intervortex potential.  Of these, the former is usually
much larger, as noted above, and has already been evaluated in
eq.\ (6).  We substitute this value into eq.\ (3) and hence into 
the expression for $E_{WC}$.  In the limit $a_0 \ll \lambda_{\perp}$, 
one can evaluate this sum numerically.  The result is very well fitted
numerically by the form
$\sum_{\bf l \neq 0}\K\Bigl({l\over\lp}\Bigr) \approx n_v\int{d^2{\bf r}
\K\Bigl(
{r\over\lp}\Bigr)}-0.500{\rm ln}\Bigl(\lp^2 n_v\Bigr) -1.437$.
Collecting all these results, we finally obtain 
\begin{equation}
{E_{WC}\over{2\epsilon_0 dS}}={{n_v}^2\over 2}\int{d^2{\bf r}\K\Bigl(
{r\over\lp}\Bigr)}- 
0.25\ n_v\ {\rm ln}\Bigl(\lp^2 n_v\Bigr) - 
\end{equation}
$$
-0.719\ n_v+{n_v^2\over 2 n_p}.
$$

For the liquid phase, the wave function symmetry matters
since the pancakes are delocalized.  We use as a trial wave function
an (unnormalized) wave function of the Laughlin form\cite{laughlin}:
\begin{equation}
\Psi_{LL,m}= 
\prod_{j < k}(z_j-z_k)^m\exp(-\frac{1}{4}\sum_{\ell}|z_{\ell}|^2).
\end{equation}
Here $z_j = x_j+iy_j$ is the position coordinate of the j$^{th}$ pancake, 
and all lengths are expressed in units of the 
``magnetic length'' $\ell_0 \equiv (\Phi_0/(2\pi B_{eff}))^{1/2}$. 
Since the vortex pancakes are bosons, $m$ must be an even integer.  In the
Laughlin theory of the fractional quantum Hall effect, $1/m$ is the filling
fraction of the first Landau level.

Laughlin's prescription for obtaining
the minimizing value of $m$ is readily translated to the present problem, 
in which the role of charges and magnetic field are reversed.   
The generalized prescription is that the minimizing $m$ 
occurs when the number density n$_p$ of vortices of the fictitious
magnetic field equals $m$ times the number density $n_v$ of 
fictitious charges, i.\ e. $m = n_p/n_v$.
  
We next calculate the internal energy of the Laughlin liquid
at various values of $m$.  With a change of scale, the 
vortex-vortex interaction energy of the liquid becomes
$E_{LL}/(2\epsilon_0 dS)=(n_v/2\pi)\int{d^2{\bf x}\K\Bigl(
{x\over\lp\sqrt{\pi n_v}}\Bigr)g(x)}$,
where 
$g(x)$ is the dimensionless
density-density correlation function for the Laughlin liquid (normalized to
unity at large $x$),  
and ${\bf x}$ is a dimensionless coordinate defined by
${\bf x}={\bf r}\sqrt{\pi n_v}$. 
Since $g(x)$ differs significantly from unity mainly in the region $x<1$, it
is convenient to decompose the interaction energy as follows:
\begin{equation}
{E_{LL}\over{2\epsilon_0 dS}}={n_v^2\over 2}\int{d^2{\bf r}\K\Bigl(
{r\over\lp}\Bigr)}+
\end{equation}
$$
{n_v\over 2\pi}\int{d^2{\bf x} \K\Bigl(
{x\over\lp\sqrt{\pi n_v}}\Bigr)\Bigl(g(x)-1\Bigr)}\approx
$$
$$
{n_v^2\over 2}\int{d^2{\bf r}\K\Bigl({r\over\lp}\Bigr)}-
$$
$$
-n_v\int_0^{\infty}{xdx\Biggl({\rm ln}\Bigl({x\over 2\lp\sqrt{\pi n_v}}\Bigr)+
\gamma\Biggr)\Bigl(g(x)-1\Bigr)},
$$
where $\gamma \approx 0.577...$ is Euler's constant and we have used the 
small-$x$ approximation for K$_0$(x).

As noted by Laughlin, the correlation function g(r) for 
the Laughlin liquid state is just that of the 
2D one-component classical plasma (OCP), in which the
particles interact logarithmically.  The last term on the right is, 
to within a factor, just the internal energy of the OCP.    
We can therefore use standard numerical results for the OCP, as obtained
by Monte Carlo methods by Caillol {\it et al}\cite{levesque}. 
Using the analytical fit of these authors to their own numerical results for
the integral $\int_0^{\infty}{xdx{\rm ln}\/x(g(x)-1)}$, we find
\begin{equation}
-\int_0^{\infty}{xdx\Biggl({\rm ln}\Bigl({x\over 2\lp\sqrt{\pi n_v}}\Bigr)+
\gamma\Biggr)\Bigl(g(x)-1\Bigr)}=
\end{equation}
$$
-\int_0^{\infty}{xdx{\rm ln}\/x\Bigl(g(x)-1\Bigr)-
{1\over4}{\rm ln}\Bigl(4\pi\lp^2 n_v\Bigr)+{\gamma\over2}}\approx
$$
$$
-0.3755+0.4400\Bigl({n_v\over2n_p}\Bigr)^{0.74}-{1\over4}{\rm ln}
\Bigl(\lp^2 n_v\Bigr)-{1\over4}{\rm ln}(4\pi)+{\gamma\over2}.
$$
Hence, the energy of the Laughlin liquid can be written as
\begin{equation}
{E_{LL}\over{2\epsilon_0 dS}}={{n_v}^2\over 2}\int{d^2{\bf r}\K\Bigl(
{r\over\lp}\Bigr)}-{1\over 4}n_v\ {\rm ln}\Bigl(\lp^2 n_v\Bigr) - 
\end{equation}
$$
0.720\ n_v+0.4400\ n_v\Bigl({n_v\over 2 n_p}\Bigr)^{0.74}
$$

Finally, the zero-temperature melting transition is defined by the 
equation $E_{WC}=E_{LL}$, or
${n_v\over2 n_p}\approx0.440\Bigl({n_v\over2 n_p}\Bigr)^{0.74}$,
or equivalently
\begin{equation}
{n_v\over n_p}\approx 0.09.
\end{equation}
We see that this result agrees remarkably
well with the Lindemann criterion.

\section{Discussion.}

We now evaluate these predictions for two materials of interest, using
a simplified approximation for $n_p$.  As noted
by Ao and Thouless, $n_p$ is not simply the areal density of Cooper 
pairs per unit area, but that of \underline{superconducting}
Cooper pairs - that is, those not pinned
by lattice disorder.  Since it is not obvious how to evaluate this
quantity, we simply use the London equation to estimate
$n_p$ at zero field.
To get $n_p(B)$, we use the 
Ginzburg-Landau approximation $\lambda(B,0)=\lambda(0,0)/[1-B/B_{c2}]^{1/2}$,
where $B_{c2}$ is the $T=0$ upper critical field, and $\lambda$(B, T) is the
magnetic field and temperature dependent penetration depth.  The melting
condition, from either the Lindemann criterion or from
equating solid and liquid energies, is $n_v/n_p = \beta$, where 
$\beta \approx 0.1$.  Substituting the above expressions into
this melting condition, we obtain 
\begin{equation}
\frac{B_m}{B_{c2}}= \frac{B_0}{B_0+B_{c2}},
\end{equation}
where 
$B_0 = \beta m_pc^2d\Phi_0/[4\pi\lambda^2(0,0)q^2]$.

First, we apply this result to an amorphous MoGe film, an extensively studied
2D extreme Type-II
superconductor.  An amorphous Mo$_{0.43}$Ge$_{0.57}$ film
of thickness $30 \AA$ has  
$\lambda(0,0) \approx 8000\AA$ and $B_{c2} \approx 10^4$ G\cite{ephron}.
Taking $B \approx H$ (a good approximation in the extreme Type-II limit),
and using $\beta = 0.1$, we find 
$B_0 \approx 7 \times 10^4$ G, and therefore $B_m/B_{c2} \approx 0.8-0.9$.
This is consistent with the observations of Ephron {\it et al}\cite{ephron},
who find a superconducting-insulating transition at 
around 10 kG, quite close to the estimated $B_{c2}$.  The transition
in \cite{ephron} is undoubtedly \underline{not} uncomplicated quantum
melting, since it occurs in highly disordered samples.  
Indeed, it is undoubtedly better described as a continuous phase transition 
from
a vortex glass to a Cooper pair glass\cite{fisher}.  Nonetheless, it is 
gratifying that our predicted field, estimated for a \underline{clean} sample, 
falls rather close to the observed transition.

Of at least equal interest is possible quantum melting
in high-T$_c$ superconductors.  Since our model is strictly
2D, we consider only a single layer of a high-T$_c$ material.  The result
may conceivably be extrapolated to the most anisotropic CuO$_2$-based
high-T$_c$ materials, such as BiSr$_2$Ca$_2$Cu$_2$O$_{8+x}$.  Assuming
$d = 10 \AA$ and  $\lambda(0,0) = 1400 \AA$, we obtain 
$B_0 \approx 1.5\times 10^6$ G.  Estimating $B_{c2} = 3 \times 10^6$ G,
we find $B_m \approx 10^6$ G.  Since $T_c$ is smaller and
$\lambda(0,0)$ is larger in an underdoped sample, however,
we may expect $B_m$ also to decrease in such materials. 
   
Finally, we comment on the connection between our results and the 
calculations of \cite{onogi}.  While these authors find FQHE-like
commensuration effects in the flux liquid state, their observed melting scales
with m$_v$ as if there were no influence of
$B_{eff}$ on $B_m$.  Our simplified analytical calculations suggest that 
$B_{eff}$ may dominate the melting behavior for sufficiently light pancake 
masses ($m_v \ll 40m_e$).  Presumably, this influence of $B_{eff}$ 
would show up in QMC studies at sufficiently low values of m$_v$.

To conclude, we have calculated the quantum 
melting criterion for a 2D vortex lattice at $T = 0$, by comparing the
internal energies of the vortex solid and vortex fluid states in a hypothetical
superclean limit.  We find that, at sufficiently low vortex masses,
melting behavior seems to be dominated by a fictitious magnetic field 
acting on the vortices and produced by the Cooper pair density.  
The calculated melting field is
close to the superconducting-insulating transition observed in certain 
thin films of amorphous $MoGe$, and may
be within reach of pulsed magnetic fields in some underdoped 
CuO$_2$-based high-T$_c$ materials.

\section{Acknowledgments.}

One of us (DS) gratefully acknowledges many valuable conversations with
Professor S.\ Doniach, as well as the warm hospitality of the 
Department of Applied Physics at Stanford University, where this calculation
was initiated.  This work was supported by NSF Grant DMR94-02131 and
by the Department of Energy 
through the Midwest Superconductivity Consortium at Purdue
University, through Grant DE-FG90-02ER-45427.

\end{document}